\documentclass[%
 reprint,
 amsmath,amssymb,
 aps,
]{revtex4-1}

\usepackage{graphicx}
\usepackage{dcolumn}
\usepackage{bm}
\usepackage{epstopdf}

\begin{document}


\title{Calibration and Performance of the ATLAS Tile Calorimeter}%

\author{Bernardo Sotto-Maior Peralva, on behalf of the ATLAS Collaboration}
\affiliation{Federal University of Juiz de Fora, Brazil}


\date{\today}

\begin{abstract}
The Tile Calorimeter (TileCal) is the hadronic calorimeter covering the most central region of the ATLAS experiment at the LHC. It is a key detector for the measurement of hadrons, jets, tau leptons and missing transverse energy. The TileCal calibration system comprises radioactive source, laser and charge injection elements and it allows to monitor and equalize the calorimeter response at each stage of the signal production, from scintillation light to digitization. This contribution presents a brief description of the different TileCal calibration systems as well as the latest results on their performance in terms of calibration factors, linearity and stability. The performance of the Tile Calorimeter with the cosmic muons and collision data is also presented, including the absolute energy scale, time resolution and associated stabilities.
\end{abstract}

\maketitle


\section{\label{sec:intro} The ATLAS Tile Calorimeter}

\begin{figure}[!h]
\centering
\begin{center}
  \includegraphics[scale=.4]{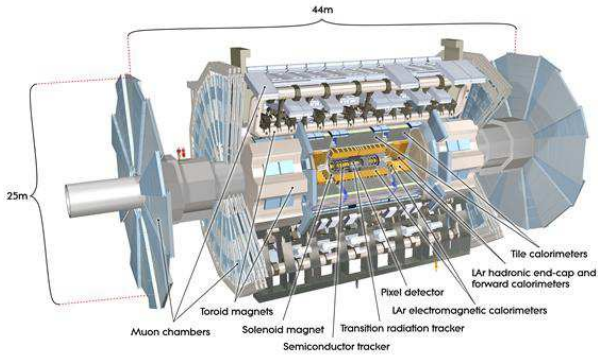}
  \caption{The ATLAS detector and its subsystems.}\label{fig:atlas}
\end{center}
\end{figure}

The ATLAS (A Toroidal LHC ApparatuS) \cite{atlas} is one of the four main experiments at the LHC and one of two general purpose detectors designed for precision Standard Model measurements and to search for physics beyond the Standard Model. It is composed of six different subsystems: The Inner Detector, the Solenoidal Magnet that surrounds the inner detector, the Electromagnetic and Hadronic calorimeters, the Toroid Magnets and the Muon Spectrometer, as illustrated in Figure~\ref{fig:atlas}.

\begin{figure}[!h]
\centering
\begin{center}
  \includegraphics[scale=.3]{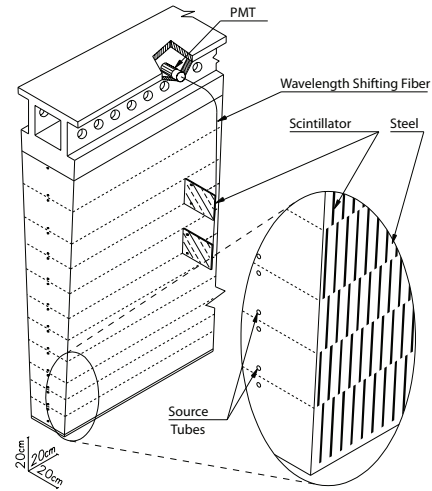}
  \caption{Schematic showing the mechanical assembly and the optical "readout of the Tile Calorimeter, corresponding to a $\phi$ wedge.}\label{fig:tile}
\end{center}
\end{figure}

The Tile Calorimeter (TileCal) \cite{tile} is the hadronic calorimeter of ATLAS. It consists of three cylinders, one long barrel (LB) splitted into two readout partitions, LBA and LBC, and two extended barrels (EBs), EBA and EBC, covering the most central region $|\eta|< 1.7$ of the ATLAS detector. The focus of TileCal is to perform precise measurements of hadrons, jets, taus and the missing transverse energy as well as to provide input signal to the Level 1 Calorimeter Trigger. TileCal is a sampling device which uses iron plates as absorber and plastic scintillating tiles as the active material. The particles produced in the interaction point travel through the calorimeter and the light produced in the scintillating tiles is proportional to the energy deposited by the particles. The light is transmitted by wavelength shifting fibers and read out by photomultiplier tubes (PMTs) which generate analog pulses. Figure~\ref{fig:tile} shows the structure and the signal collection system of one of the 256~$\phi$ wedges of TileCal.

The collected signals are amplified and shaped in such a way that the amplitude is proportional to the energy~\cite{frontend}. The analog pulses are received by the digitizers where they are sampled every 25~ns. Finally, the signal samples are stored for the digital processing phase where amplitude and phase of the pulse are reconstructed.~\cite{ofv}.

\section{\label{sec:tilecal}The TileCal Calibration Systems}

The calibration systems are important to finely adjust the energy in a channel by channel basis as well as to identify potential hardware failures. The final cell ener\-gy is computed by taking into account the following constants:

\begin{equation}\label{eq:calib}
E[GeV]=A[ADC] \cdot C_{ADC\rightarrow pC} \cdot C_{laser} \cdot C_{Cs} \cdot C_{pC\rightarrow GeV}
\end{equation}
where $A[ADC]$ is the pulse amplitude in ADC counts.

The last constant $C_{pC\rightarrow GeV}$ is a factor to convert charge to electromagnetic scale energy and was determined from the test beam measurements in 2001-2003 using electron beams~\cite{calib}. Other calibration constants are defined using dedicated calibration systems which monitor and calibrate the front-end readout circuit as well as the PMTs and the optics. These systems are described in the following sections.

\subsection{\label{sec:cis} The Charge Injection System}

The Charge Injection System (CIS)~\cite{frontend} is used to derive ADC to pC conversion factors for the digital readout. The CIS simulates physics signals in TileCal channels by generating pulses from discharge capacitors in the readout circuit and measuring the electronic response. Two capacitors of 5.2~pF and 100~pF are used and the pulse amplitude is controlled by a 10-bit ADC. It provides then a quantitative relationship between the analog signals from the Tile Calorimeter photomultiplier tubes and the electronic response of TileCal readout channels. A set of CIS calibration constants are regularly produced and applied to TileCal data. The precision achieved by the CIS is within 0.7\%.

\subsection{\label{sec:cesium} Cesium System}

The Cesium (Cs) system~\cite{cesium} is based on a movable radioactive source using hydraulic control. A $^{137}$Cs $\gamma$-source traverses through all the tiles in order to equalize the response between individual channels and to monitor stability of optics elements. The channel response to the energy deposited is used to equalize the response of all the cells and maintain global response of the calorimeter at the electromagnetic scale. The Cesium calibration tests the optical chain including the scintillators ageing effect.
Cs scans are performed every one, two months. The precision of the determinations is 0.3\%.

\subsection{\label{sec:laser} Laser System}

The gain of each PMT is measured using a Laser calibration system~\cite{laser} that sends a controlled amount of light from a laser (532 nm, 10 ps pulse) to monitoring photodiodes and each of the TileCal PMT simultaneously. The stability of the diodes is monitored and a set of filters allows to adapt the light intensity. The Laser measurements are used for fast monitoring of TileCal, for timing calibration and to correct the PMTs gain variations between two Cs scans. The typical precision of the laser system is better than 0.5\% on the gain variation determination.

\subsection{\label{sec:integrator} The Integrator System}
In the high energy proton-proton collisions at the LHC the dominating processes are soft parton interactions, or Minimum Bias~(MB) events~\cite{minbias}. The integrator system of each PMT integrates the response over time and it is used to measure the average signal of the MB interactions during proton-proton collisions. The MB measurements allow a determination of the luminosity. We are working to use the MB results in the calibration chain of the detector.

\section{\label{sec:operation} Operation of the Tile Calorimeter}

\begin{figure}[!h]
\centering
\begin{center}
  \includegraphics[scale=.3]{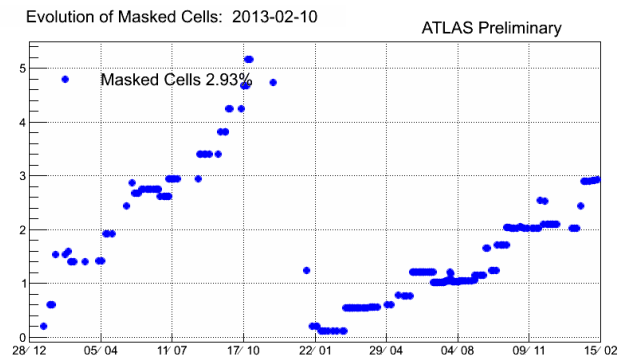}
  \caption{Time evolution of the fraction of unusable TileCel cells.}\label{fig:maskTimeLine}
\end{center}
\end{figure}

\begin{figure}[!h]
\centering
\begin{center}
  \includegraphics[scale=.35]{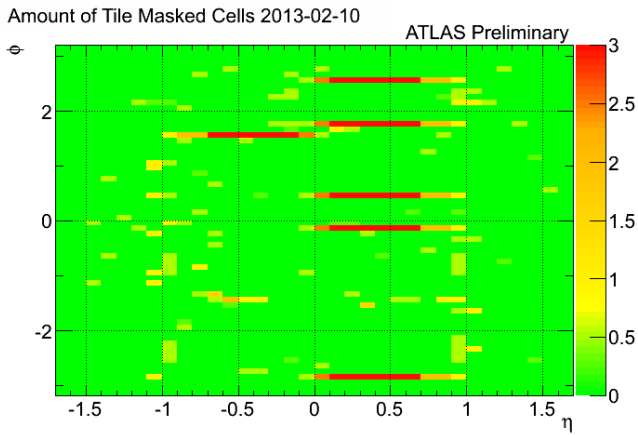}
  \caption{TileCal masked cells at the beginning of February 2013 (switched-off modules are highlighted).}\label{fig:maskEtaPhi}
\end{center}
\end{figure}

In the year of 2011, TileCal provided 99.2\% of good data for physics. As shown in Figure~\ref{fig:maskTimeLine}, a fraction of the TileCal cells (up to 5\% at the end of 2011) were masked. Most of the masked channels came from modules that were off due to Low Voltage Power Supply (LVPS) problems. The LVPSs are installed on the front-end electronics located in the detector hall, in a high radiation environment. Figure~\ref{fig:maskTimeLine} shows the time evolution of the fraction of unusable cells starting from the beginning of 2011 until February 2013. It can be noticed that du\-ring the 2011-12 winter shutdown, almost all cells were re\-covered. At the end of Period 1 (Middle of February 2013) TileCal had approximately 3\% of its cells masked and, once again, the main reason was the fact that the corresponding modules were off due to LVPS problems, as can be seen in Figure~\ref{fig:maskEtaPhi}.

The LVPS has been re-designed in order to become robust against the power trips observed during the data taking. At the beginning of 2012 the revised version of the LVPS was installed in 40 modules. In addition, the new LVPS reduces the non-gaussian component of the noise distribution, thus resulting in a global reduction of the noise values (see the discussion in the next section).

\subsection{\label{sec:noise} Electronic noise}

\begin{figure}[!h]
\centering
\begin{center}
  \includegraphics[scale=.4]{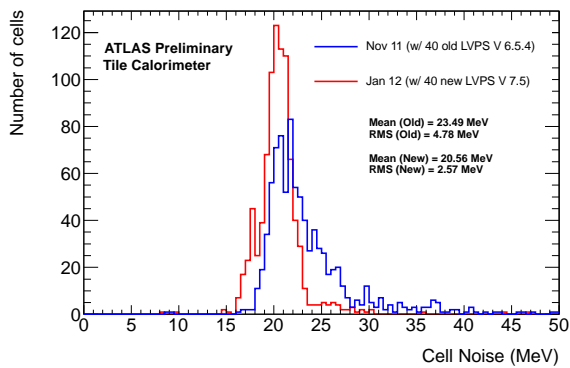}
  \caption{Comparison between the noise with new and old LVPS.}\label{fig:cellNoise}
\end{center}
\end{figure}

TileCal exploits dedicated runs without the presence of collisions to measure the electronic noise. Noise parameters are used to set trigger thresholds and design energy estimation filters. In 2011, using the old LVPS, the electronic noise deviates from single gaussian mainly due to the instability of the old LVPS. With the new LVPS, the non-gaussian component was reduced as shown in Figure~\ref{fig:cellNoise} where the rms of the signal distributions are reported.

\section{\label{performance} Performance}

\subsection{\label{sec:combined} Combined calibration}

When combining results from all calibration systems, one can determine what causes changes in the overall detector response. In the first two years of operation an updrift in the PMT gains was observed. This effect disappeared in 2011 when the
LHC started operating at higher luminosity than previously. Since then, a downdrift is observed during the time the beam is on, and a slow recovery when the beam is off. This effect is seen by the Cesium scans, the integrator system and the laser calibrations. Since all three subsystems show similar behavior, we conclude that it is caused mainly by drift of the PMT gains. It is mostly affecting PMTs belonging to cells at lower radius which receive more scintillator light.

\begin{figure}[!h]
\centering
\begin{center}
  \includegraphics[scale=.55]{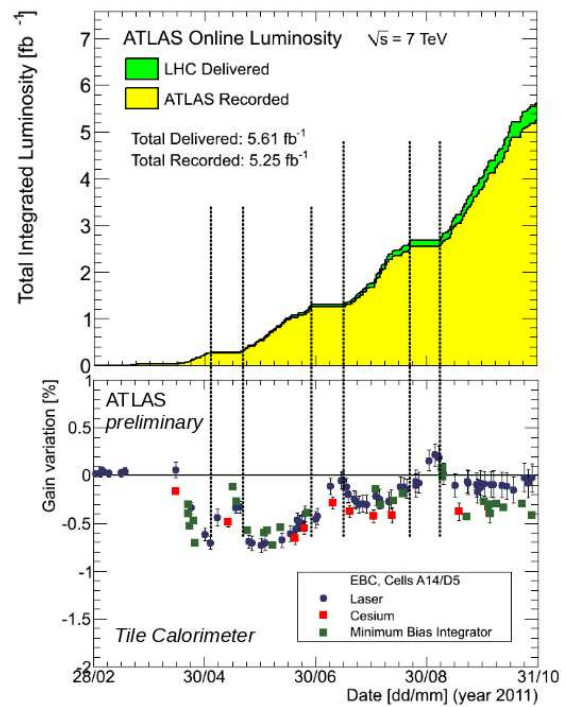}
  \caption{Gain variation measured by three calibration systems, for one TileCal cell (bottom), compared to LHC integrated luminosity (top) in 2011.}\label{fig:lumi}
\end{center}
\end{figure}

Figure~\ref{fig:lumi} shows the evolution of ATLAS total integrated luminosity (top) and the evolution of the
response of a cell at low radius (bottom). The response is measured separately by the Cesium, Laser and
Minimum Bias calibration systems.

\subsection{\label{sec:mc} Checks of the calibration using cosmic rays data}
A fraction of 11\% of the TileCal modules were cali\-brated in the beam tests in 2001-2003. Electron and muon beams were used to establish the electromagnetic scale (EM) and to inter-calibrate different radial layers. After installation of the whole TileCal in the ATLAS experimental hall, cell inter-calibration was done with the help of the Cesium calibration system. The Cesium
source was moved through every calorimeter cell and the high voltage of every PMT was adjusted to have the cell response equal to the response measured during beam tests. The comparison between cosmic ray data and Monte Carlo (MC) prediction and between beam test muons and cosmic muons has confirmed that propagation of the electromagnetic scale from the beam tests to ATLAS was successful. Non-uniformity within one layer as seen by muons turned out to be at the level of 2-3\%, the maximal difference between layers is 4\%.

\subsection{\label{sec:isolatedHadrons} $E/p$ from isolated hadrons}

\begin{figure}[!]
\centering
\begin{center}
  \includegraphics[scale=.66]{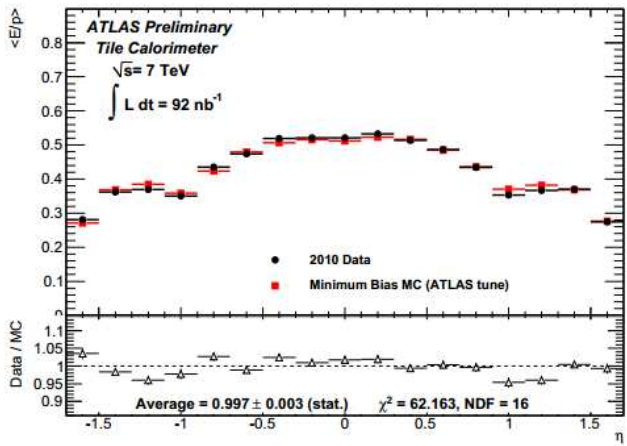}
  \caption{Mean value of the $E/p$ as a function of $\eta$, integrated over all $p$ ranges for 2010 collision data and Monte Carlo simulation. Bottom plot shows ratio of data to Monte Carlo of the mean value of the $E/p$. The value shown is a fitted value with a constant function. The error shown is only statistical.}\label{fig:isolatedHadrons}
\end{center}
\end{figure}

The calorimeter energy response is also investigated using isolated charged hadrons. Isolated tracks having
energy deposits compatible with minimum ionizing particles in the electromagnetic calorimeter, which is in front of TileCal, are selected. Their energy, $E$, deposited in TileCal is compared with momentum, $p$, measured in the tracking detectors, giving $E/p$. Figure~\ref{fig:isolatedHadrons} shows the mean values of $E/p$ as a function of $\eta$ for the $\sqrt{s} = 7$ TeV $pp$ collision data and MC simulation. The data is reasonably described by Monte Carlo within $\pm5\%$.

\vspace{1cm}

\section{\label{conclusions} Conclusions}

The calibration systems of the ATLAS TileCal have been presented. The CIS, laser and cesium calibration systems allow to monitor the evolution of the response of the different components of TileCal with a 0.5-1.0\% precision. Analysis of the combined calibrations can be used to gain detailed insight in what causes variations in detector response. Monte Carlo simulation agrees with data for response to muons and single hadrons. During the phase-I shutdown all the old LVPS will be replaced with the new version.

\begin{acknowledgments}
I would like to thank the TileCal group for the fruitful discussion, and also the UFJF, CNPq, CAPES and FAPEMIG for the financial support.
\end{acknowledgments}


\begin{thebibliography}{90}




%
%
%
%

\bibitem {atlas} ATLAS Collaboration, \textit{The ATLAS Experiment at the CERN Large Hadron Collider}, JINST 3 (2008) S08003.

\bibitem{tile} G. Aad, et al., \textit{Readiness of the ATLAS Tile Calorimeter for LHC collisions}, The European Physical Journal
C - Particles and Fields 70 (2010) 11931236.

\bibitem {frontend} K. Anderson et al., \textit{Design of the front-end analog electronics for the ATLAS tile calorimeter}, Nucl. Instrum. Meth. \textbf{A} 551 (2005) 469.

\bibitem {ofv} G. Usai et al., \textit{Signal Reconstruction of the ATLAS Hadronic Tile Calorimeter: implementation and
performance}, 2011, J. Phys.: Conf. Ser. 293 012056.

\bibitem {calib} P. Adragna et al., \textit{Testbeam studies of production modules of the ATLAS tile calorimeter}, Nucl. Instrum. Meth. \textbf{A} 606 (2009) 362.

\bibitem {cesium} E. Starchenko et al., \textit{Cesium monitoring system for ATLAS Tile Hadron Calorimeter}, Nucl. Instrum. Meth. \textbf{A} 494 (2002) 381.
    
\bibitem {laser} S. Viret [LPC ATLAS Collaboration], \textit{LASER monitoring system for the ATLAS tile calorimeter}, Nucl.
Instrum. Meth. \textbf{A} 617 (2010) 120.

\bibitem {minbias} ATLAS Collaboration, \textit{Charged-particle multiplicities in pp interactions at $\sqrt{s} = 900$ GeV measured with the ATLAS detector at the LHC}, Phys Lett \textbf{B} 688 (2010) 21.

\end{thebibliography}
\end{document}